# Thickness-Dependent Morphology and Electrical Transport in Ultrathin Amorphous MoSi Films

O.O. Leha[1], O. V. Zraichenko[1], S. I. Kryvonohov[2], V.Yu. Lyakhno[1]

*[1]B. Verkin Institute for Low Temperature Physics and Engineering of the National Academy of Sciences of Ukraine, 47 Nauky Ave., Kharkiv, 61103, Ukraine*

*[2]Institute for Single Crystals, STC "Institute for Single Crystals", National Academy of Sciences of Ukraine, 60 Nauky Ave., Kharkiv, 61001, Ukraine*

## Abstract

We investigate the thickness-dependent surface morphology and electrical transport of ultrathin amorphous $Mo_{50}Si_{50}$ films deposited by magnetron co-sputtering. Atomic force microscopy of Si-capped heterostructures on $Si/SiO_2$ reveals smooth surfaces throughout the investigated thickness range, with RMS roughness below 0.2 nm and a non-monotonic thickness dependence characterized by a minimum at 1.92 nm. Room-temperature sheet resistance follows an approximately inverse-thickness dependence for films with nominal thicknesses of about 1.9-5 nm, whereas a pronounced deviation from this scaling develops below approximately 1.5-2 nm. Temperature-dependent measurements further reveal a superconducting transition in the 30-nm film, while the 1.54-nm film shows a strong resistance increase upon cooling clearly indicates a superconductor-insulator transition. In the 2-20 K range, the sheet conductance of the 1.54-nm film exhibits a reproducible logarithmic temperature dependence, $G_s(T)=G_0+B\ln(T/T_0)$, consistent with quantum corrections to diffusive transport in a strongly disordered quasi-two-dimensional conductor. These results identify a thickness-sensitive transport regime in ultrathin MoSi while demonstrating that the corresponding heterostructures retain comparatively smooth surface morphology.

*Keywords: amorphous MoSi, ultrathin films, thin-film growth, atomic force microscopy, sheet resistance, superconducting devices*

## 1. Introduction

Superconductivity in structurally disordered materials has been studied since the work of Anderson, who showed that strong nonmagnetic elastic scattering is compatible with conventional (s)-wave superconductivity [1]. Experimental studies subsequently established superconductivity in amorphous transition-metal alloy films [2]. In amorphous Mo-Si systems, superconductivity and high upper critical fields were demonstrated in sputter-deposited alloys [3], while correlations between normal-state transport and superconducting properties were investigated over different alloy compositions [4]. Subsequent studies further characterized the superconducting properties of amorphous MoSi-based films [5], while binary molybdenum compounds more broadly have been considered promising materials for both fundamental superconductivity studies and device applications [6]. More recent work has established amorphous MoSi as a particularly attractive thin-film superconductor because it combines structural homogeneity, relatively high normal-state resistivity, and superconductivity in layers only a few nanometers thick [7].

These properties make MoSi particularly attractive for superconducting nanowire single-photon detectors (SNSPDs) and related quantum-photonic devices. The first SNSPDs were demonstrated using NbN nanowires [8]. Subsequently, single-photon detection was demonstrated using 4-nm-thick amorphous MoSi films [9], and later MoSi SNSPDs achieved high detection efficiency together with low timing jitter [10]. Integration of amorphous MoSi nanowire detectors with nanophotonic waveguides has also been demonstrated [11]. More broadly, SNSPDs have developed into an important detector platform for quantum communication, sensing, imaging, and integrated quantum photonics [12].

Film thickness becomes a critical parameter when the MoSi layer is reduced to only a few nanometers. Structural studies have revealed short-range atomic correlations below approximately 1 nm and medium-range ordering on scales of about 1-3 nm in amorphous MoSi [7]. In addition, in-situ studies of sputter-deposited Mo-Si films have demonstrated that the amorphous-to-crystalline transformation can depend strongly on both composition and deposited thickness [13]. In films with thicknesses of approximately 1-5 nm, these characteristic structural length scales become comparable to the total film thickness, motivating a detailed investigation of whether morphology, electrical transport, and superconducting properties continue to follow simple thickness scaling in the ultrathin limit.

Several basic thin-film relations illustrate why this regime is physically important. For a homogeneous conducting layer of thickness ($d$), the sheet resistance is related to the resistivity by $R_s = \rho/d$. Thus, if the intrinsic resistivity remains approximately unchanged, the product $R_s d$ should also remain approximately constant. The reduction in thickness additionally modifies characteristic superconducting length scales. In the two-dimensional limit $d << \lambda$, where $\lambda$ is the magnetic penetration depth, magnetic screening is described by the Pearl length $\Lambda = 2\lambda^2/d$, which increases strongly as the film becomes thinner [14]. The Pearl model has also been experimentally applied to vortex and Meissner-screening measurements in amorphous MoSi thin films [15]. For a disordered superconducting film in the local dirty limit and at low temperature, the sheet kinetic inductance can be approximated by $L_{k,\square}=\hbar R_s/\pi\Delta_0$, where $\Delta_0$ is the superconducting energy gap [16]. Within weak-coupling BCS theory $\Delta_0 \approx 1.764k_BT_c$, so that $L_{k,\square} = \hbar R_s/1.764\pi\ k_BT_c$[17]. For strongly disordered superconductors, however, deviations from the conventional Mattis-Bardeen response associated with a broadened superconducting density of states can modify this simple kinetic-inductance estimate [18].

The strong dependence of sheet resistance and kinetic inductance on film thickness is directly relevant to superconducting detector and resonator technologies. High-kinetic-inductance disordered superconducting alloys have been investigated for kinetic-inductance detectors [19], while microwave kinetic-inductance detectors exploit the kinetic response of superconducting charge carriers for sensitive radiation detection [20]. Control of ultrathin-film transport properties is therefore important not only for SNSPDs but also for a broader class of superconducting electromagnetic and quantum devices.

At the same time, the simple $R_s{\sim}1/d$ scaling may break down in sufficiently thin disordered films. In low-dimensional disordered conductors, deviations from conventional Boltzmann transport can arise from Anderson localization and electron–electron interaction effects [21]. In superconducting films, increasing disorder can additionally suppress superconductivity even when the film remains structurally homogeneous [22]. Consequently, a pronounced increase in sheet resistance in the ultrathin limit does not by itself imply granular or physically discontinuous film growth.

In the present work, we investigate the early growth stages of amorphous $Mo_{50}Si_{50}$ films, with particular emphasis on the approximately 1-5 nm thickness range. Atomic force microscopy is used to characterize the thickness-dependent surface morphology of Si-capped heterostructures, while four-point measurements are employed to study sheet resistance and its deviation from simple geometrical scaling. Temperature-dependent resistance measurements provide additional information on the evolution of the transport regime. The aim is to correlate nanoscale morphology with electrical transport and to identify characteristic thickness regimes relevant to the formation and application of ultrathin amorphous MoSi films.

## 2. Methods

MoSi films were deposited by direct-current magnetron co-sputtering from separate molybdenum and silicon targets. Three types of substrates were used: sapphire $Al_2O_3$, thermally oxidized silicon Si/$SiO_2$ and silicon coated with silicon nitride Si/$Si_3N_4$. The Si/SiO2 substrates contained a thermally grown $SiO_2$ layer with a thickness of 237 nm. Prior to deposition, the substrates were prepared as 10x10 $mm^2$ samples. The present analysis focuses on films deposited on Si/$SiO_2$ substrates, which are commonly used in SNSPD fabrication.

To remove residual particles and surface contamination introduced during substrate preparation and handling, the samples were sequentially cleaned in acetone, isopropyl alcohol, and deionized water using an ultrasonic bath. The cleaned substrates were then loaded into a high-vacuum magnetron sputtering system. The base pressure in the deposition chamber prior to argon admission was approximately $2x10^{-6}$ mbar, while the argon working pressure during sputtering was maintained at $2x10^{-3}$ mbar. No intentional substrate heating was applied.

Mo and Si were sputtered simultaneously from separate targets, allowing the power applied to each magnetron to be controlled independently. The deposition rate was controlled by the corresponding magnetron powers. For $P_{Si}$ = 101 W and $P_{Mo}$=64 W, the calibrated deposition rates were 6.6 nm/min for Si and 5 nm/min for Mo, respectively, corresponding to a total nominal growth rate of approximately 11.6 nm/min. For $P_{Si}$=125 W and $P_{Mo}$ =81 W, the respective deposition rates were 8.1 and 6.3 nm/min, giving a total nominal growth rate of approximately 14.4 nm/min.

During deposition, the substrate holders were continuously rotated to average spatial variations in the particle flux from the individual targets and thereby improve the lateral uniformity of both film thickness and composition across the substrate surface. Films with a nominal Mo:Si atomic ratio of 50:50 and different MoSi thicknesses were investigated. For the 0.96- and 1.54-nm films, the sputtering powers were $P_{Si}$=101 W and $P_{Mo}$=64 W, corresponding to a calibrated total deposition rate of approximately 11.6 nm/min. For the 1.92-, 2.4-, and 5-nm films, the sputtering powers were increased to $P_{Si}$=125 W and $P_{Mo}$=81 W, corresponding to a total nominal deposition rate of approximately 14.4 nm/min. The nominal film thicknesses were determined from the calibrated deposition rates. The uncertainty in the nominal film thickness was estimated from the uncertainty of the deposition process and was taken into account in the thickness-dependent analysis, where it is represented by horizontal error bars in Figures 4 and 5. To protect the ultrathin MoSi layer from oxidation and environmental exposure, all films were incorporated into multilayer structures consisting of a 5.4-nm Si underlayer, the MoSi layer of variable thickness, and a 2.7-nm Si capping layer.

Previous studies have demonstrated amorphous structure in ultrathin sputter-deposited MoSi films, including 4-nm films used for superconducting single-photon detectors [9]. In-situ structural studies of Mo-Si additionally show that the amorphous-to-crystalline transformation depends on film composition and thickness [13]. Taken together, these results support the expectation that the nominal $Mo_{50}Si_{50}$ films investigated here remain in the amorphous regime under the employed room-temperature sputter-deposition conditions. No independent structural determination of the amorphous state was performed for the present samples due to their low thickness.

Surface morphology was characterized by atomic force microscopy (AFM) over approximately 0.9x0.9 $\mu m^2$ areas using identical spatial sampling for all samples. In addition to the Si/MoSi/Si heterostructures, two reference surfaces were analyzed under the same conditions: the bare Si/$SiO_2$ substrate and a Si/$SiO_2$ substrate covered by an 8.1-nm Si layer, corresponding to the total Si thickness used as the 5.4-nm underlayer and 2.7-nm capping layer in the MoSi heterostructures. The AFM maps were processed using mean-plane subtraction followed by fourth-order polynomial background correction, and the surface parameters $S_q$, $S_a$, and $S_z$ were determined from the corrected topographies, where $S_q$ is the root-mean-square height deviation from the mean surface plane, $S_a$ is the arithmetic mean of the absolute height deviations, and $S_z$ is the total peak-to-valley height range. Similar AFM-based surface-morphology analyses have been applied in previous thin-film studies [23-25].

Room-temperature electrical characterization was performed using a collinear four-point probe configuration. The measured resistance values were converted to sheet resistance $R_s$ using the appropriate geometrical correction factor for the sample dimensions and probe geometry, following the standard four-point-probe approach for thin conducting layers [26]. Measurements were performed for films deposited on $Al_2O_3$, Si/$SiO_2$, and Si/$Si_3N_4$ substrates.

For the analysis of the temperature-dependent measurements, the measured four-terminal resistance $R(T)$ was converted to sheet resistance using a temperature-independent geometrical factor determined from the room-temperature measurements of the same sample. The conversion factor was obtained by matching the measured resistance at 300 K to the independently determined room-temperature sheet resistance $R_s(T)=R(T)\times[R_s(300,K)/R(300,K)]$. The corresponding sheet conductance was calculated as $G_s(T)=1/R_s(T)$. This procedure assumes that the sample geometry and contact configuration remain unchanged throughout the temperature sweep.

## 3. Results

### 3.1. Thickness-dependent surface morphology

Atomic force microscopy was used to investigate the surface morphology of Si-capped MoSi heterostructures with nominal MoSi thicknesses of 0.96, 1.92, 2.4, and 5.0 nm. Since the MoSi layer was covered by a 2.7-nm Si capping layer, the measured topography represents the surface response of the complete heterostructure rather than the exposed MoSi layer itself. Approximately 0.9x0.9 $\mu m^2$ areas with identical spatial sampling were used for the quantitative comparison. Representative larger-area AFM topographies are shown in Figure 1.

Following mean-plane subtraction and fourth-order polynomial background correction, a pronounced non-monotonic dependence of surface roughness on MoSi thickness was observed (Table 1). The 0.96-nm structure exhibited an RMS roughness of $S_q$=0.177 nm, with $S_a$=0.139 nm and $S_z$=1.445 nm. Increasing the MoSi thickness to 1.92 nm resulted in substantial surface smoothing, with $S_q$ decreasing to 0.098 nm, corresponding to a reduction of approximately 44%. The 1.92-nm sample therefore exhibited the smoothest surface within the investigated series.

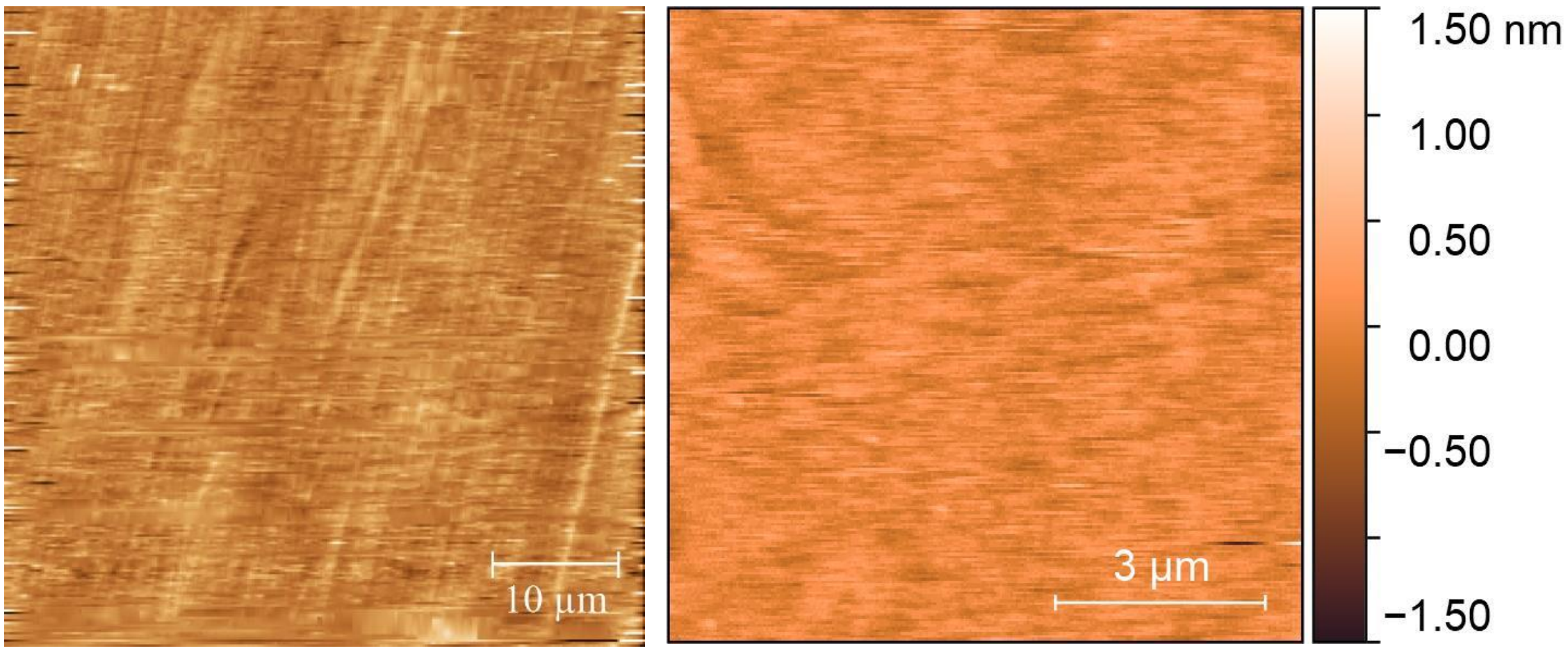

Figure 1. AFM topography of the bare $Si/SiO_2$ substrate (50x50 $\mu m^2$, left) and the Si/MoSi/Si heterostructure with a nominal MoSi thickness of 0.96 nm (right). The color scale indicates surface height.

A further increase in thickness to 2.4 nm led to a pronounced reversal of this trend. The RMS roughness increased to $S_q$=0.193 nm, while $S_a$ and $S_z$ increased to 0.152 and 1.668 nm, respectively. Thus, the RMS roughness at 2.4 nm was almost twice that of the 1.92-nm structure. At 5 nm, the surface became smoother again, with $S_q$=0.124 nm, $S_a$=0.099 nm, and $S_z$=1.028 nm.

| Structure | Nominal MoSi thickness, nm | $S_q$, nm | $S_a$, nm | $S_z$, nm |
|---|---|---|---|---|
| substrate ($Si/SiO_2$) | - | 0.262 | 0.206 | 3.085 |
| substrate ($Si/SiO_2$)/8.1-nm Si | - | 0.360 | 0.272 | 5.240 |
| Si/MoSi/Si | 0.96 | 0.177 | 0.139 | 1.445 |
| Si/MoSi/Si | 1.92 | 0.098 | 0.079 | 0.798 |
| Si/MoSi/Si | 2.40 | 0.193 | 0.152 | 1.668 |
| Si/MoSi/Si | 5.00 | 0.124 | 0.099 | 1.028 |

Table 1. AFM surface roughness parameters of the bare $Si/SiO_2$ substrate, the 8.1-nm Si reference layer, and Si-capped MoSi heterostructures with different nominal MoSi thicknesses on $Si/SiO_2$ substate.

Thus, the surface morphology does not evolve through simple monotonic smoothing or roughening. Instead, a distinct roughness minimum occurs at 1.92 nm, followed by pronounced roughening at 2.4 nm and subsequent smoothing at 5 nm. For comparison, the bare $Si/SiO_2$ substrate exhibited $S_q$=0.262 nm, $S_a$=0.206 nm, and $S_z$=3.085 nm, whereas the 8.1-nm Si reference layer yielded $S_q$=0.360 nm, $S_a$=0.272 nm, and $S_z$=5.240 nm. Thus, all investigated Si/MoSi/Si heterostructures exhibited lower RMS roughness than both reference surfaces. Even the 2.40-nm structure, which represents the roughness maximum within the MoSi thickness series, remained substantially smoother than the bare substrate and the Si-only reference layer.

### 3.2. Surface height-profile analysis

Representative vertical height profiles were extracted from the processed AFM maps near the center of each image. The analyzed profile length was approximately 902 nm for all samples, providing an additional means of comparing the local vertical and lateral characteristics of the surface morphology (Table 2).

| $d$, nm | $R_a$, nm | $R_q$, nm | $R_t$, nm | $R_{sk}$ | $R_{ku}$ | $S_m$, nm | $\lambda_q$, nm | $\Delta_q$ |
|---|---|---|---|---|---|---|---|---|
| 0.96 | 0.143 | 0.188 | 1.107 | 0.236 | 3.494 | 17.69 | 18.57 | 0.0636 |
| 1.92 | 0.075 | 0.092 | 0.531 | 0.094 | 2.916 | 16.40 | 18.83 | 0.0308 |
| 2.40 | 0.122 | 0.161 | 1.041 | 0.115 | 4.014 | 15.03 | 16.92 | 0.0598 |
| 5.00 | 0.086 | 0.108 | 0.632 | −0.089 | 3.049 | 18.41 | 19.61 | 0.0347 |

Table 2. Roughness, spatial, and hybrid parameters obtained from representative AFM height profiles of Si-capped MoSi heterostructures deposited on Si/$SiO_2$ substrates at different nominal MoSi thicknesses. Here, $d$ is the nominal MoSi thickness; $R_a$ is the arithmetical mean roughness of the profile; $R_q$ is the root-mean-square (RMS) profile roughness; $R_t$ is the total peak-to-valley height of the roughness profile; $R_{sk}$ is the profile skewness coefficient; $R_{ku}$ is the profile kurtosis; $S_m$ is the mean spacing between profile irregularities; $\lambda_q$ is the RMS wavelength of the profile, characterizing the lateral scale of surface variations; and $\Delta_q$ is the RMS profile slope.

The profile analysis reproduces the non-monotonic trend obtained from the areal AFM parameters (Figure 2). Between 0.96 and 1.92 nm, $R_q$ decreases from 0.188 to 0.092 nm and $R_t$ from 1.107 to 0.531 nm, corresponding to decreases of approximately 51% and 52%, respectively. At the same time, $\lambda_q$ remains almost unchanged, changing only from 18.57 to 18.83 nm, whereas the RMS slope $\Delta_q$ decreases from 0.0636 to 0.0308. Thus, the transition from 0.96 to 1.92 nm is dominated by a reduction in the vertical amplitude and local steepness of the surface features rather than by a substantial change in their characteristic lateral scale.

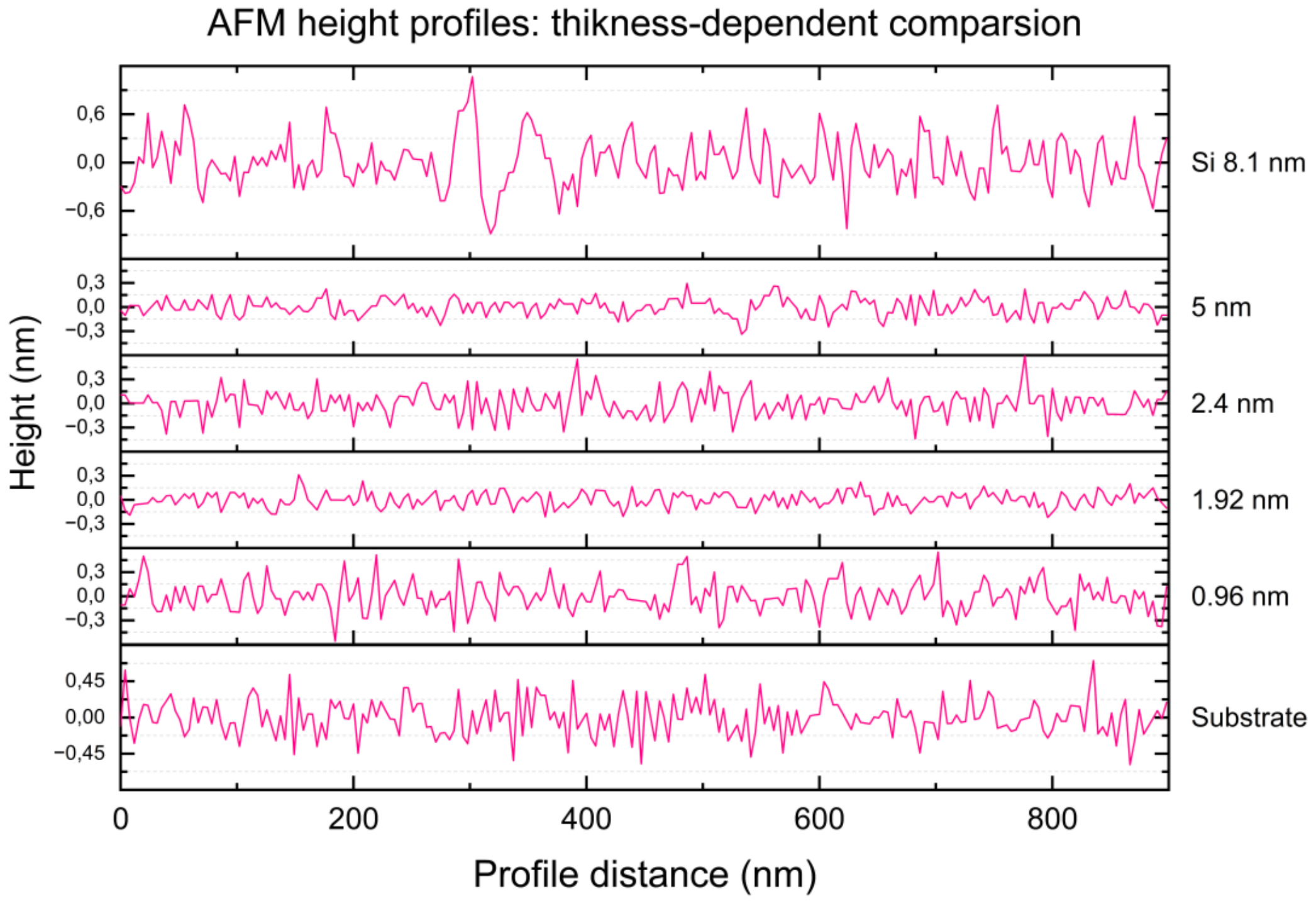

Figure 2. Representative AFM height profiles of the bare Si/$SiO_2$ substrate, the 8.1-nm Si reference layer, and Si-capped MoSi heterostructures with nominal MoSi thicknesses of 0.96, 1.92, 2.4, and 5.0 nm. All profiles were extracted over comparable distances and are shown to illustrate the evolution of the local surface-height variations across the reference and MoSi-containing structures.

At 2.4 nm, $R_q$ increases to 0.16 nm and $R_t$ to 1.04 nm. The RMS slope simultaneously increases to 0.0598, while $\lambda_q$ decreases to 16.92 nm. The height distribution also changes: $R_{ku}$ increases from 2.92 at 1.92 nm to 4.01 at 2.4 nm, indicating a greater contribution from pronounced local height excursions.

At 5 nm, $R_q$ decreases again to 0.11 nm and $R_t$ to 0.63 nm. The RMS slope decreases to 0.035, whereas $\lambda_q$ increases to 19.61 nm. The kurtosis returns to $R_{ku}$=3.05, close to the value expected for an approximately Gaussian height distribution.

The profile analysis therefore confirms the sequence identified by the areal AFM measurements. The 2.4-nm structure is distinguished not only by enhanced vertical roughness but also by steeper local features, a shorter characteristic lateral scale, and a larger contribution from extreme height deviations.

### 3.3. Room-temperature electrical transport

The room-temperature sheet resistance of MoSi films deposited on Si/$SiO_2$ decreases strongly with increasing nominal film thickness. The measured $R_s$ values are summarized in Table 3.

| Nominal d, nm | Ω/sq |
|---|---|
| 0.96 | 18518 |
| 1.54 | 3589 |
| 1.92 | 2241 |
| 2.40 | 2269 |
| 2.40 | 1990 |
| 2.88 | 1543 |
| 4.08 | 696 |
| 5.00 | 855 |
| 30.0 | 125 |

Table 3. Room-temperature sheet resistance of $Mo_{50}Si_{50}$ films deposited on Si/$SiO_2$ as a function of nominal thickness.

The dominant trend is a pronounced reduction in sheet resistance as the MoSi thickness increases (Figure 3). In the ultrathin range, $R_s$ decreases from $1.85\times10^4$ Ω/sq at 0.96 nm to 3589 Ω/sq at 1.54 nm and 2241 Ω/sq at 1.92 nm.

Two nominally 2.4-nm samples yielded sheet resistances of 2269 and 1990 Ω/sq, corresponding to a difference of approximately 14%. Since the film thickness was determined from calibrated deposition parameters rather than by an independent thickness measurement of each individual sample, part of this variation may reflect uncertainty in the actual deposited thickness. This uncertainty was therefore retained in the thickness-dependent analysis and is represented by the horizontal error bars in Figures 4 and 5.

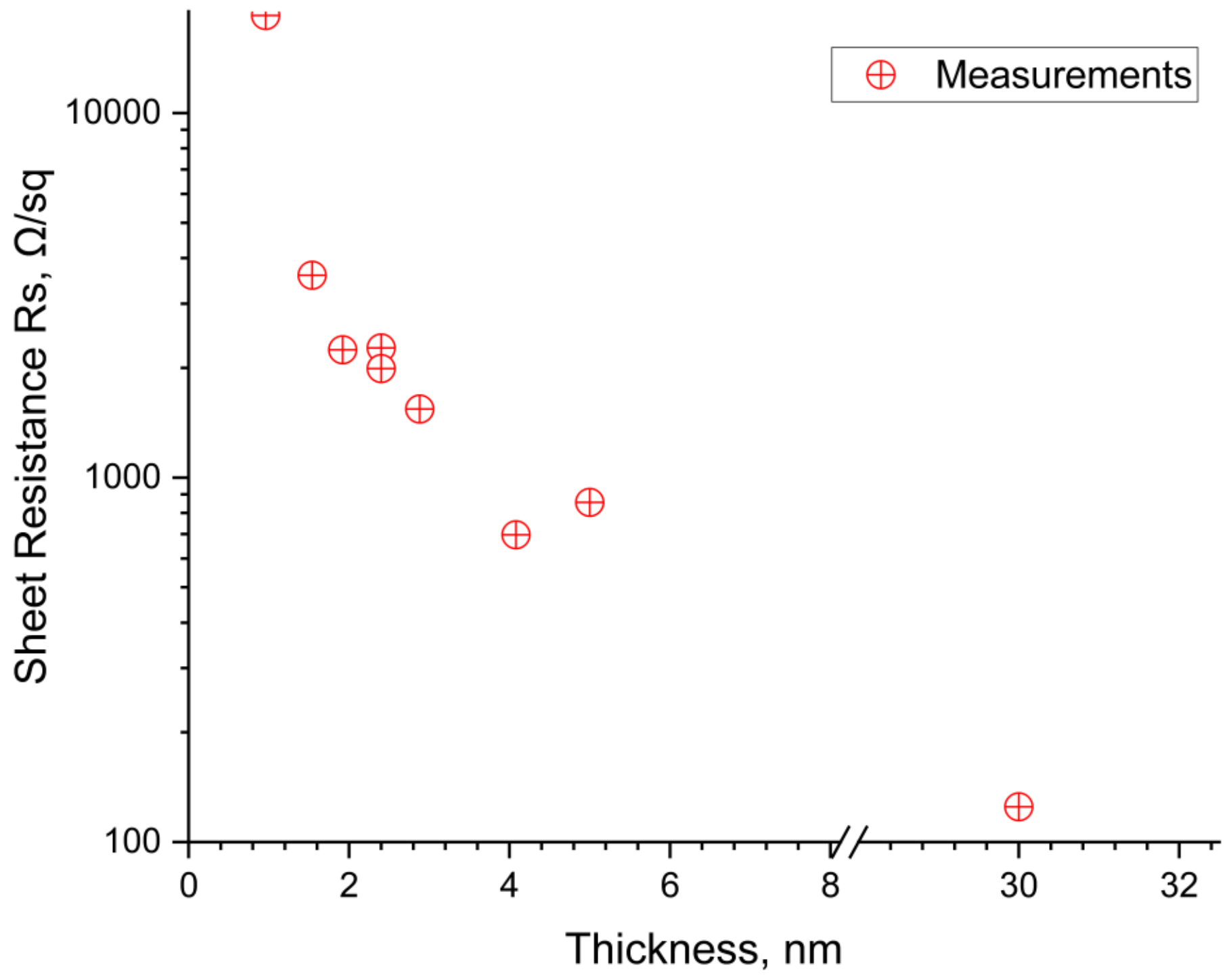

Figure 3. Thickness dependence of sheet resistance in $Mo_{50}Si_{50}$ thin films deposited on $Si/SiO_2$ substrates.

To evaluate the thickness dependence quantitatively, the data in the approximately 1.9–5 nm range were compared with a power-law relation $R_s = Ad^{-n}$. Using the 1.92-, 2.40-, 2.88-, and 5.00-nm data points, with $R_s$=1990 Ω/sq for the additional 2.40-nm sample, gives $n = 1.04$ with $R^2 = 0.988$. Thus, $R_s \sim d^{-1.04} \approx d^{-1}$, which is close to the inverse-thickness dependence expected for a laterally continuous film with approximately thickness-independent resistivity. A closely related near-unity scaling exponent has been reported by Erbe et al. for MoSi films deposited on Si. Using the universal scaling relation $T_c d = AR_s^{-B}$, they obtained $B$=1.06±0.11, corresponding to a sheet-resistance exponent statistically consistent with unity [27]. Although this relation simultaneously involves $T_c$, $d$, and $R_s$ and is therefore not identical to the present room-temperature $R_s(d)$ analysis, the near-unity exponent is consistent with an approximately inverse scaling of sheet resistance with film thickness in this thickness regime. The 4.08-nm point shows a pronounced deviation from this trend and was therefore not included in the power-law fit. Since no independent thickness verification is presently available for this sample, this deviation should be treated as a sample-specific experimental variation rather than as evidence for an additional transport regime.

In contrast, the strong increase in sheet resistance below approximately 1.5-2 nm cannot be described by the same inverse-thickness scaling, indicating the onset of a thickness-sensitive transport regime.

### 3.4. Effective resistivity parameter

To separate the geometrical contribution to sheet resistance from changes in the intrinsic transport properties of the ultrathin layer, the product $\rho_{eff} = R_s{*}d$ was evaluated. For a homogeneous continuous film, $\rho_{eff}$ corresponds to the electrical resistivity. For the thinnest films, where the assumptions of a homogeneous bulk-like conductor may no longer be fully valid, it is referred to here as an effective resistivity-like parameter (Table 4).

| Nominal (d), nm | $R_s$, Ω/sq | $R_s{*}d$, μΩ·cm |
|---|---|---|
| 0.96 | 18518 | 1778 |
| 1.54 | 3589 | 553 |
| 1.92 | 2241 | 430 |
| 2.40 | 2269 | 545 |
| 2.40 | 1990 | 478 |
| 2.88 | 1543 | 444 |
| 4.08 | 696 | 284 |
| 5.00 | 855 | 428 |
| 30.00 | 125 | 375 |

Table 4. Nominal thickness $d$, measured sheet resistance $R_s$, and calculated effective resistivity-like parameter $R_s{*}d$ for $Mo_{50}Si_{50}$ thin films deposited on $Si/SiO_2$ substrates.

For the 1.92-, 2.4-, 2.88-, and 5-nm films used in the inverse-thickness analysis, the $\rho_{eff}$ values lie predominantly within approximately 428-478 μΩ·cm with a mean value of approximately 445 μΩ·cm and a median value 437 μΩ·cm. The near-constant $\rho_{eff}$ values in this thickness range provide an alternative representation of the same result obtained from the power-law fit: the sheet resistance is approximately governed by $R_s \approx \rho_{eff}/d$. A qualitatively different behavior is observed in the thinnest films. At 1.54 nm, $\rho_{eff} \approx 553$ μΩ·cm, whereas at 0.96 nm it increases sharply to $\rho_{eff} \approx 1778$ μΩ·cm. Thus, the strong increase in $R_s$ near the 1-nm limit cannot be explained solely by the reduction of the conducting cross section. Instead, the effective transport properties of the film themselves become strongly thickness dependent. The combined $R_s(d)$ and $\rho_{eff}(d)$ dependences therefore indicate two characteristic regimes: an approximately inverse-thickness regime for $d \gtrsim 1.9$ nm and a strongly thickness-dependent transport regime below approximately 1.5-2 nm as illustrated on Figures 4 and 5.

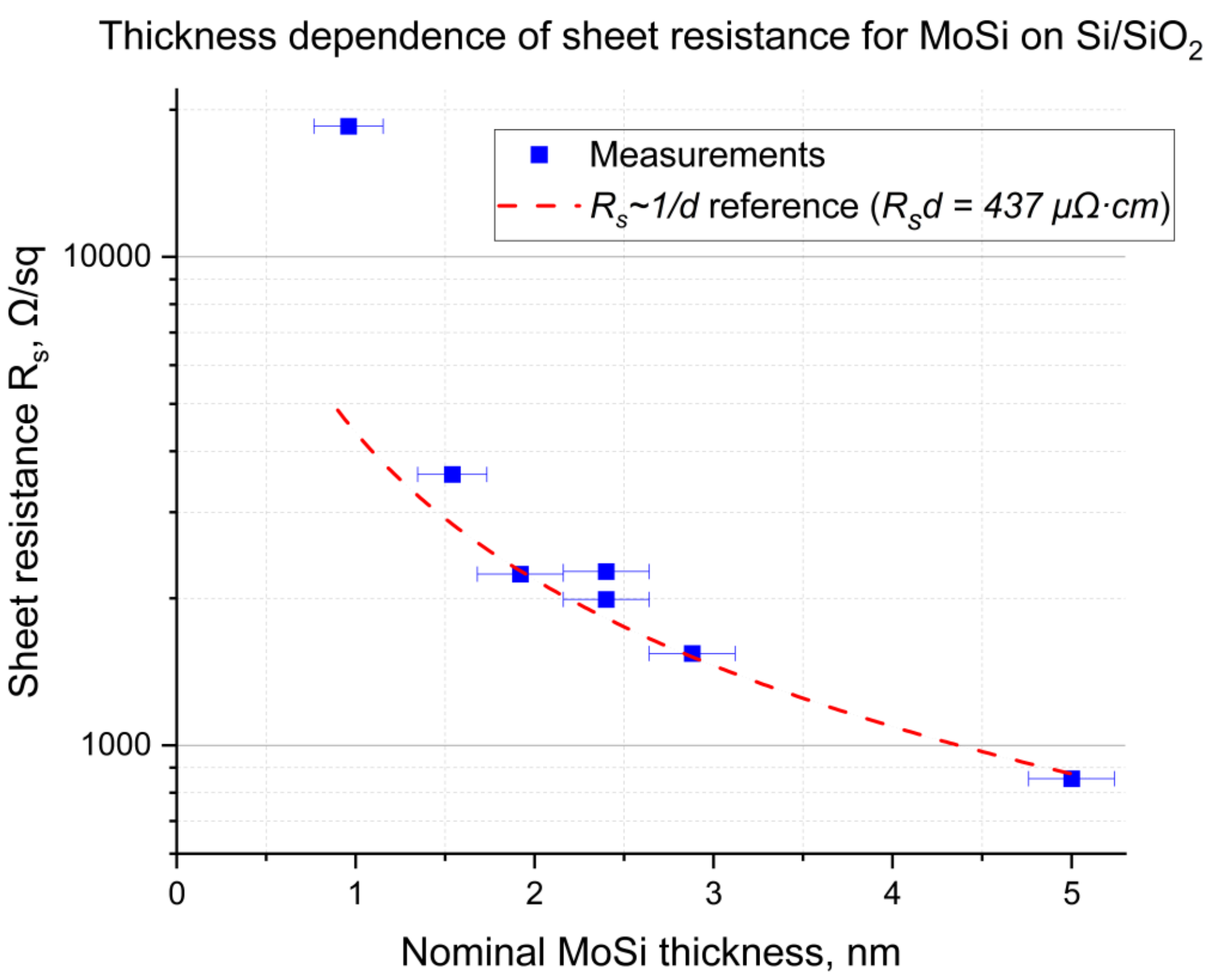

Figure 4. Room-temperature sheet resistance $R_s$ as a function of nominal MoSi thickness for films deposited on $Si/SiO_2$. The dashed line shows the $R_s \sim d^{-1}$ reference behavior for $R_s * d$=437 μΩ·cm. Horizontal error bars represent the estimated uncertainty in the nominal film thickness associated with the deposition process.

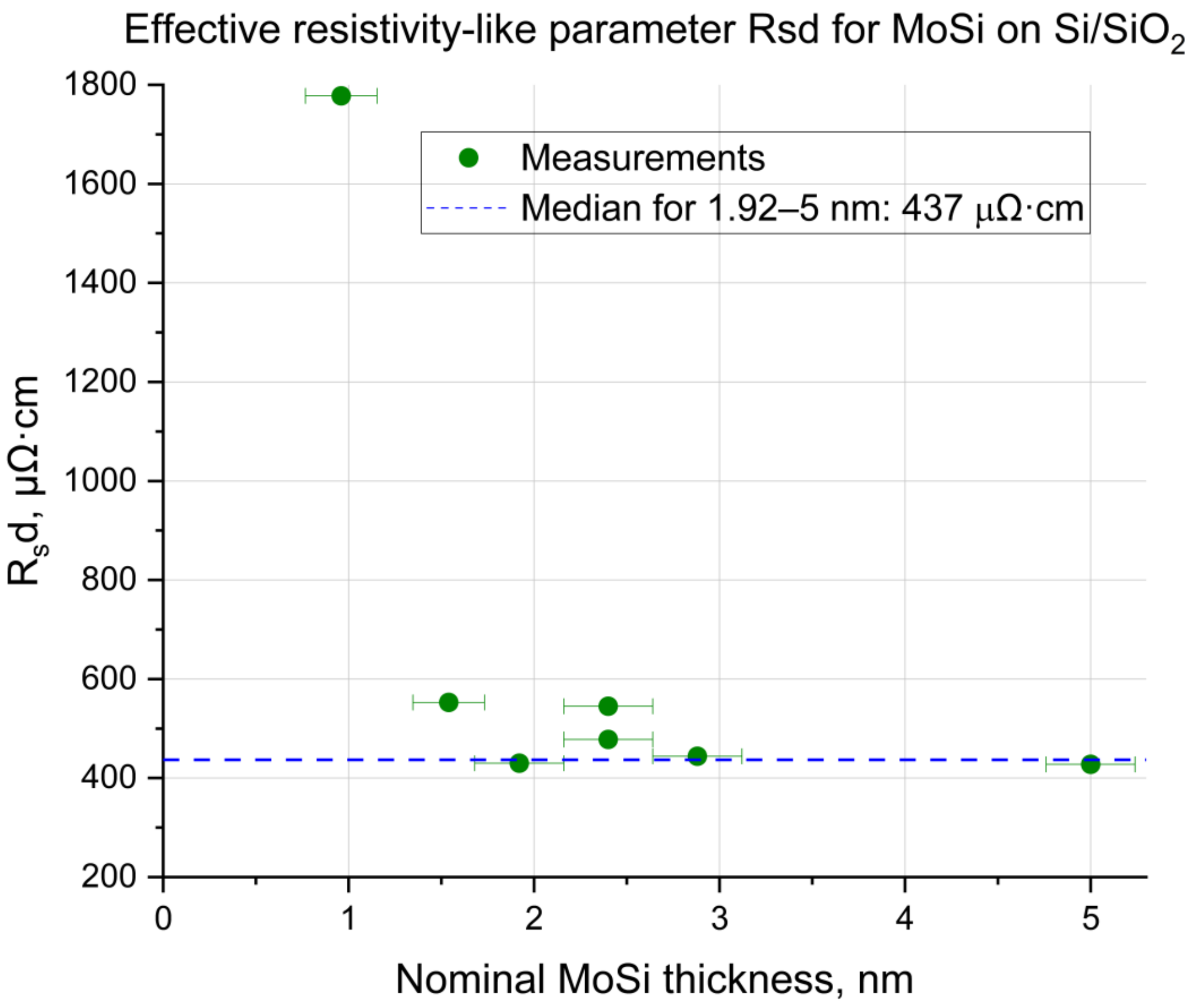

Figure 5. Effective resistivity-like parameter $\rho_{eff}$ as a function of nominal MoSi thickness for films deposited on $Si/SiO_2$ substrates. The dashed horizontal line indicates the median $R_s{*}d$ value of 437 μΩ·cm for the 1.92-5 nm thickness range. Horizontal error bars represent the estimated uncertainty in the nominal film thickness associated with the deposition process.

### 3.5. Temperature-dependent resistance

Temperature-dependent resistance measurements reveal a pronounced difference between the thick- and ultrathin-film limits (Figure 6). To compare the shapes of the temperature dependences independently of their absolute resistance values, the data were normalized to the room-temperature resistance, $R_{norm}(T)=R(T)/R(300\ K)$. The absolute room-temperature sheet resistances of the 1.54- and 30-nm films deposited on $Si/SiO_2$ were 3589 Ω/sq and 125 Ω/sq, respectively, corresponding to a difference of nearly a factor of 29. The 30-nm film exhibits only a weak variation of the normalized resistance over most of the normal-state temperature range, followed by a sharp decrease at low temperature associated with the superconducting transition. In contrast, the 1.54-nm film shows a pronounced negative temperature coefficient of resistance, $dR/dT < 0$ with the resistance increasing continuously upon cooling. At the lowest temperatures shown, the normalized resistance reaches approximately $R/R300$~1.8, corresponding to an increase of about 80% relative to its room-temperature value. No superconducting transition is observed for the 1.54-nm film within the measured temperature range. The markedly different low-temperature responses demonstrate that reducing the MoSi thickness from 30 to approximately 1.5 nm produces not only a large increase in room-temperature sheet resistance but also a qualitative change in the character of electrical transport.

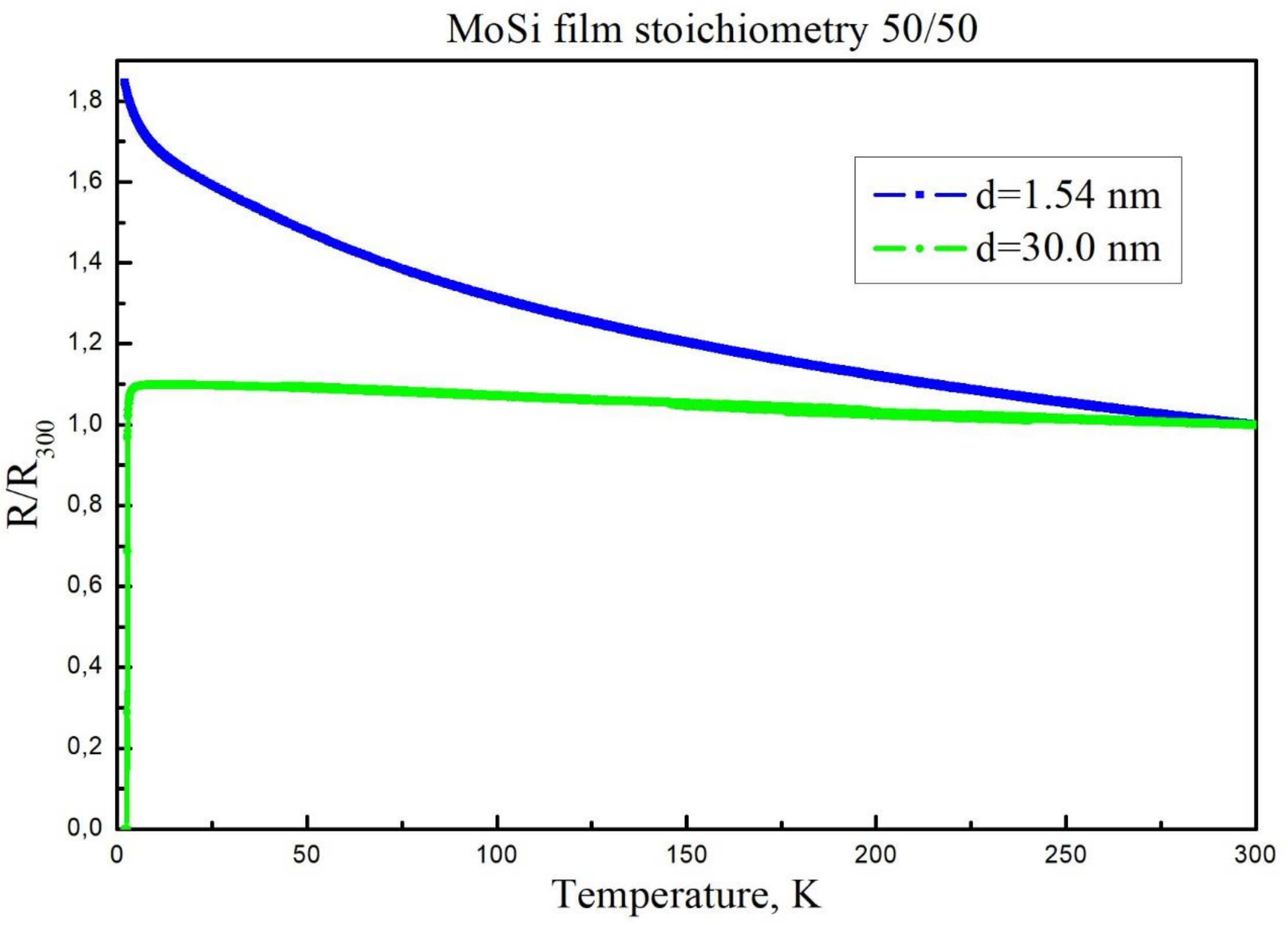

Figure 6. Temperature dependence of the normalized resistance $R/R_{300}$ for $Mo_{50}Si_{50}$ films with nominal thicknesses of 1.54 and 30 nm. The two samples exhibit qualitatively different temperature-dependent transport behavior.

## 4. Discussion

The combined morphological and electrical measurements reveal a pronounced thickness-dependent evolution of ultrathin MoSi films. For films with nominal thicknesses in the approximately 1.9–5 nm range, the sheet resistance generally follows the inverse-thickness dependence expected for a continuous conducting layer, $R_s = \rho/d$. A power-law analysis of representative data in this thickness range yields an exponent close to unity, $R_s \sim d^{-1.04}$, indicating that the dominant thickness dependence is approximately consistent with geometrical scaling. This conclusion is further supported by the effective resistivity-like parameter $\rho_{eff} = R_s{*}d$, which remains close to 440 μΩ·cm for most films in this thickness range. Since $R_s{*}d$ can be interpreted as the material resistivity only for a homogeneous continuous film, it is treated here as an effective transport parameter rather than as a strictly intrinsic resistivity.

A qualitatively different behavior appears as the nominal MoSi thickness is reduced below approximately 1.5-2 nm. For the 1.54-nm film, $R_s{*}d$ increases to approximately 553 μΩ·cm, while the 0.96-nm film reaches approximately 1778 μΩ·cm. Thus, the rapid increase in sheet resistance of the thinnest films cannot be attributed solely to the reduction of the conducting cross section. Instead, the data indicate a progressively stronger influence of disorder, reduced dimensionality, interface-related scattering, and possible localization effects on electrical transport.

Similar thickness-dependent electronic behavior has been observed in other homogeneous amorphous superconducting films. Graybeal and Beasley reported localization-related transport in ultrathin amorphous MoGe films that remained morphologically homogeneous down to approximately 2 nm, together with a systematic suppression of $T_c$ as the sheet resistance increased [28]. In MoSi itself, Banerjee et al. obtained a fitted critical thickness of $d_c$=1.46±0.15 nm, below which superconductivity was expected to disappear within their model [7]. Erbe et al. subsequently obtained $d_c$=1.38±0.03 nm for MoSi films on Si and 1.87±0.10 nm for films on GaAs using the Simonin thickness-scaling model [27]. Although the compositions, substrates, and deposition conditions differ between these studies and the present samples, these characteristic thicknesses fall within the same approximately 1.5–2 nm range in which the present $R_s(d)$ data begin to deviate strongly from inverse-thickness scaling.

The temperature-dependent resistance measurements independently support this crossover. The 30-nm MoSi film exhibits only a relatively weak increase in normal-state resistance upon cooling and undergoes a sharp superconducting transition at approximately 2.6-2.7 K. In contrast, the 1.54-nm film shows a pronounced resistance increase with decreasing temperature and no superconducting transition down to approximately 2 K. The close agreement between cooling and warming measurements demonstrates that these temperature dependences are reproducible and that no pronounced thermal hysteresis is present. The absence of an observed transition in the 1.54-nm film is qualitatively consistent with the broader disorder-driven suppression of superconductivity expected in highly resistive homogeneous films [22, 28].

To investigate the origin of the resistance upturn in the 1.54-nm film, the low-temperature data were analyzed in terms of several transport models. The strongest agreement was obtained when the sheet conductance, $G_s = 1/R_s$, was plotted as a function of logarithmic temperature (Figure 7). In the range 2-20 K, the conductance is well described by $G_s(T) = G_0 + B*ln(T/T_0)$ where $T_0$=1 K is used as a reference temperature. Linear fits give $B$ = 9.33 μS for cooling and $B$ = 9.31 μS for warming, with $R^2$=0.99917 and 0.99918, respectively. The nearly identical slopes obtained for the two temperature sweeps demonstrate that the logarithmic temperature dependence is highly reproducible.

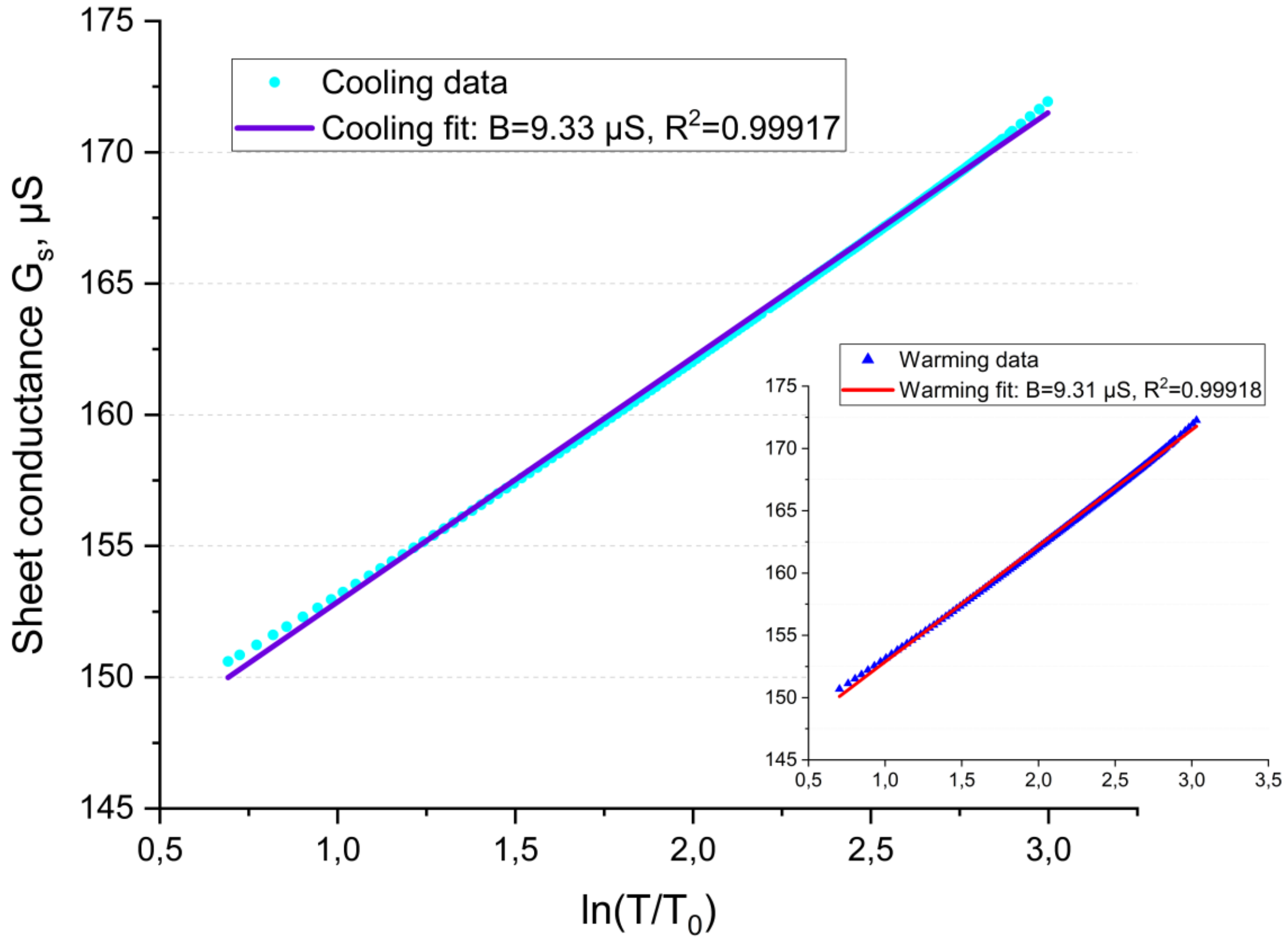

Figure 7. Sheet conductance $G_s$ of the 1.54-nm MoSi film as a function of $ln(T/T_0)$ in the low-temperature range of 2-20 K. Cooling and warming data are shown together with linear fits according to $G_s(T)=G_0+B*ln(T/T_0)$, where $T_0$=1 K. The nearly identical logarithmic slopes and high coefficients of determination for both temperature sweeps demonstrate the reproducibility of the low-temperature logarithmic conductance dependence.

Logarithmic corrections to conductivity are a characteristic feature of disordered quasi-two-dimensional conductors. Altshuler, Aronov, and Lee showed that electron-electron interactions in disordered two-dimensional Fermi systems generate logarithmic corrections to the conductivity [29]. Weak localization provides a second source of quantum corrections in thin disordered films through coherent backscattering and quantum interference [30], while the broader interplay between localization and interaction effects in disordered conductors is reviewed in Ref. [21]. The characteristic conductance scale of these two-dimensional quantum corrections is of order $e^2/(\pi h) \approx 12.33\ \mu S$. The experimentally obtained logarithmic slope of approximately 9.3 μS is therefore of the same order, corresponding to an effective dimensionless prefactor of approximately 0.75. This agreement in scale supports an interpretation of the low-temperature resistance upturn in terms of quantum corrections to diffusive transport rather than a purely geometrical increase in resistance.

Closely related quantum-correction behavior has also been investigated experimentally in Mo/Si multilayers. Buchstab et al. described the temperature dependence and magnetoresistance of such structures in terms of weak-localization, electron-electron-interaction, and superconducting-fluctuation corrections and observed a dimensional crossover in the transport behavior [31]. Although those multilayer structures are physically distinct from the amorphous MoSi alloy studied here, they demonstrate that logarithmic and interference-related corrections can be important in Mo-Si-based disordered conducting systems.

Alternative transport models were also examined for the 1.54-nm film in the 2-20 K range. This comparison is particularly relevant because Mott variable-range hopping has previously been reported for sputtered amorphous MoSi films [32]. The models tested here included two-dimensional Mott variable-range hopping, $R(T) = R_0\exp[(T_M/T)^{1/3}]$, based on hopping between localized electronic states [33], and Efros-Shklovskii hopping, $R(T)=R_0\exp[(T_{ES}/T)^{1/2}]$ which incorporates the Coulomb gap produced by interactions between localized electrons [34]. A simple thermally activated Arrhenius dependence was also examined. The corresponding linearized representations showed poorer agreement with the experimental temperature dependence than the logarithmic-conductance representation, while the characteristic temperatures extracted from the hopping fits were substantially below the investigated temperature range. The present data are therefore more consistent with logarithmic quantum corrections to diffusive transport than with the tested activated or variable-range-hopping descriptions.

The present zero-field measurements nevertheless do not allow the relative weak-localization and electron-electron-interaction contributions to be separated unambiguously. Both mechanisms can produce logarithmic temperature corrections in disordered two-dimensional conductors [21, 29, 30]. Magnetotransport measurements would provide an important additional test because a magnetic field suppresses the quantum-interference contribution associated with weak localization [30], and combined resistance and magnetoresistance measurements have previously been used to separate quantum corrections in Mo/Si systems [31].

The AFM measurements provide complementary information on the morphological evolution of the ultrathin heterostructures. All investigated Si/MoSi/Si structures exhibit low surface roughness, with $S_q$ values between 0.098 and 0.193 nm. Importantly, these values are lower than those measured for both the bare Si/$SiO_2$ substrate ($S_q$=0.262 nm) and the 8.1-nm Si reference layer ($S_q$=0.36 nm). Thus, the non-monotonic roughness evolution within the MoSi thickness series occurs entirely within a comparatively smooth surface-morphology regime.

The increase in roughness from 1.92 to 2.40 nm should therefore be interpreted as a relative change within the MoSi series rather than as progressive surface degradation. Even the 2.40-nm structure remains smoother than both reference surfaces. The incorporation of an ultrathin MoSi layer between the Si underlayer and capping layer therefore markedly modifies the final surface morphology compared with the Si-only reference and is consistent with a smooth final surface morphology of the heterostructure.

AFM probes the upper Si capping layer rather than the buried MoSi layer directly and therefore cannot independently establish microscopic continuity of the MoSi film. Furthermore, the 8.1-nm Si reference is a continuous Si layer, whereas the same total Si thickness in the heterostructures is divided into a 5.4-nm underlayer and a 2.7-nm capping layer. Nevertheless, the substantially lower roughness of the MoSi-containing structures shows that the smooth final surface cannot be explained by the total deposited Si thickness alone and indicates that incorporation of the MoSi interlayer modifies the surface evolution of the complete heterostructure.

Overall, the combined $R_s(d)$, $\rho_{eff}(d)$, and $R(T)$ results indicate two transport regimes. Films with nominal thicknesses of approximately 1.9-5 nm remain close to the inverse-thickness scaling expected for a continuous highly disordered conducting layer, whereas further reduction of the MoSi thickness produces a rapid increase in the effective resistance parameter and a pronounced low-temperature conductance correction. Together with the absence of a superconducting transition in the 1.54-nm film, the logarithmic $G_s(T)$ dependence suggests that the ultrathin regime is better described as a strongly disordered diffusive conductor with substantial quantum corrections than as a conventional activated or hopping insulator [21, 22, 28–31].

## 5. Conclusions

Ultrathin amorphous $Mo_{50}Si_{50}$ films were investigated by combining AFM and electrical transport measurements. The Si-capped heterostructures remain smooth over the studied thickness range, with RMS roughness below 0.2 nm and a minimum at 1.92 nm.

For nominal MoSi thicknesses of approximately 1.9-5 nm, the sheet resistance remains close to $R_s \sim d^{-1}$, whereas below approximately 1.5-2 nm a pronounced deviation from this scaling is observed. The 1.54-nm film also exhibits a strong resistance increase upon cooling and no superconducting transition down to about 2 K.

In the 2-20 K range, its sheet conductance follows an approximately logarithmic temperature dependence, $G_s(T)=G_0+B*ln(T/T_0)$, with nearly identical cooling and warming slopes. This behavior is more consistent with quantum corrections to diffusive transport than with the tested activated or hopping models.

Overall, the results indicate that ultrathin MoSi can retain a smooth morphology while undergoing a pronounced thickness-driven evolution of its electronic transport.

**Acknowledgements**

This work was supported by the National Research Foundation of Ukraine under Project No. 22.07/0044 and by the IEEE program "Magnetism for Ukraine 2025-2026" under Project No. 9918. The authors gratefully acknowledge M. Mikhailov for productive discussions, valuable consultations, and assistance throughout this work. We gratefully acknowledge Prof. Miroslav Grajcar and his group at Comenius University Bratislava, Slovakia, for carrying out the low-temperature measurements.

# References

1. Anderson, P. W. Theory of dirty superconductors. *Journal of Physics and Chemistry of Solids* **11**, 26–30 (1959). DOI: 10.1016/0022-3697(59)90036-8.
2. Collver, M. M.; Hammond, R. H. Superconductivity in "amorphous" transition-metal alloy films. *Physical Review Letters* **30**, 92–95 (1973). DOI: 10.1103/PhysRevLett.30.92.
3. Ikebe, M.; Muto, Y.; Ikeda, S.; Fujimori, H.; Suzuki, K. Superconductivity of bulk amorphous Mo-Si and Mo-C alloys. *Physica B+C* **107**, 387–388 (1981). DOI: 10.1016/0378-4363(81)90499-X.
4. Edelstein, A. S. et al. Correlation between the superconducting and normal state properties of amorphous molybdenum–silicon alloys. *Solid State Communications* **41**, 139–142 (1982). DOI: 10.1016/0038-1098(82)91052-3.
5. Kubo, S. Superconducting properties of amorphous MoX (X = Si, Ge) alloy films for Abrikosov vortex memory. *Journal of Applied Physics* **63**, 2033–2045 (1988). DOI: 10.1063/1.341105.
6. Shapovalov, A. P.; Belogolovskii, M. O.; Boliasova, O. O.; Kordyuk, O. A. Binary molybdenum compounds: Promising materials for novel physics of superconductivity and practical applications. *Progress in Physics of Metals* **22**, 352–381 (2021). DOI: 10.15407/ufm.22.03.352.
7. Banerjee, A. et al. Characterisation of amorphous molybdenum silicide (MoSi) superconducting thin films and nanowires. *Superconductor Science and Technology* **30**, 084010 (2017). DOI: 10.1088/1361-6668/aa76d8.
8. Gol'tsman, G. N.; Okunev, O.; Chulkova, G.; Lipatov, A.; Semenov, A.; Smirnov, K.; Voronov, B.; Dzardanov, A.; Williams, C.; Sobolewski, R. Picosecond superconducting single-photon optical detector. *Applied Physics Letters* **79**, 705–707 (2001). DOI: 10.1063/1.1388868.

9. Korneeva, Yu. P.; Mikhailov, M. Yu.; Pershin, Yu. P.; Manova, N. N.; Divochiy, A. V.; Vakhtomin, Yu. B.; Korneev, A. A.; Smirnov, K. V.; Sivakov, A. G.; Devizenko, A. Yu.; Gol'tsman, G. N. Superconducting single-photon detector made of MoSi film. *Superconductor Science and Technology* **27**, 095012 (2014). DOI: 10.1088/0953-2048/27/9/095012.
10. Caloz, M.; Perrenoud, M.; Autebert, C.; Korzh, B.; Weiss, M.; Schönenberger, C.; Warburton, R. J.; Zbinden, H.; Bussières, F. High-detection efficiency and low-timing jitter with amorphous superconducting nanowire single-photon detectors. *Applied Physics Letters* **112**, 061103 (2018). DOI: 10.1063/1.5010102.
11. Häußler, M.; Mikhailov, M. Yu.; Wolff, M. A.; Schuck, C. Amorphous superconducting nanowire single-photon detectors integrated with nanophotonic waveguides. *APL Photonics* **5**, 076106 (2020). DOI: 10.1063/5.0004677.
12. Esmaeil Zadeh, I.; Chang, J.; Los, J. W. N.; Gyger, S.; Elshaari, A. W.; Steinhauer, S.; Dorenbos, S. N.; Zwiller, V. Superconducting nanowire single-photon detectors: A perspective on evolution, state-of-the-art, future developments, and applications. *Applied Physics Letters* **118**, 190502 (2021). DOI: 10.1063/5.0045990.
13. Krause, B.; Abadias, G.; Michel, A.; Wochner, P.; Ibrahimkutty, S.; Baumbach, T. Direct Observation of the Thickness-Induced Crystallization and Stress Build-Up during Sputter-Deposition of Nanoscale Silicide Films. *ACS Applied Materials & Interfaces* **8**, 34888–34895 (2016). DOI: 10.1021/acsami.6b12413.
14. Pearl, J. Current Distribution in Superconducting Films Carrying Quantized Fluxoids. *Applied Physics Letters* **5**, 65–66 (1964). DOI: 10.1063/1.1754056.
15. Ceccarelli, L.; Vasyukov, D.; Wyss, M.; Romagnoli, G.; Rossi, N.; Moser, L.; Poggio, M. Imaging pinning and expulsion of individual superconducting vortices in amorphous MoSi thin films. *Physical Review B* **100**, 104504 (2019). DOI: 10.1103/PhysRevB.100.104504.
16. Mattis, D. C.; Bardeen, J. Theory of the Anomalous Skin Effect in Normal and Superconducting Metals. *Physical Review* **111**, 412–417 (1958). DOI: 10.1103/PhysRev.111.412.
17. Bardeen, J.; Cooper, L. N.; Schrieffer, J. R. Theory of Superconductivity. *Physical Review* **108**, 1175–1204 (1957). DOI: 10.1103/PhysRev.108.1175.
18. Baránek, M.; Neilinger, P.; Manca, D.; Turutanov, O. G.; Grajcar, M. Conductivity of strongly disordered ultra-thin MoC superconducting films investigated by transmission line spectroscopy. Physica C: Superconductivity and its Applications 635, 1354749 (2025). DOI: 10.1016/j.physc.2025.1354749.
19. Calvo, M. et al. Niobium silicon alloys for kinetic inductance detectors. *Journal of Low Temperature Physics* **176**, 518–523 (2014). DOI: 10.1007/s10909-013-1072-6.
20. Baselmans, J. J. A. Kinetic inductance detectors. *Journal of Low Temperature Physics* **167**, 292–304 (2012). DOI: 10.1007/s10909-011-0448-8.
21. Lee, P. A.; Ramakrishnan, T. V. Disordered electronic systems. *Reviews of Modern Physics* **57**, 287–337 (1985). DOI: 10.1103/RevModPhys.57.287.
22. Finkel'stein, A. M. Suppression of superconductivity in homogeneously disordered systems. *Physica B: Condensed Matter* **197**, 636–648 (1994). DOI: 10.1016/0921-4526(94)90267-4.
23. Kryvonogov, S. I.; Krukhmalev, A. A.; Nizhankovskiy, S. V.; et al. Specific features of the surface morphology of modified AlN/sapphire substrates fabricated by thermochemical nitridation. Crystallography Reports 60, 138–142 (2015). DOI: 10.1134/S1063774515010125.
24. Rogacheva, E. I.; Budnik, A. V.; Dobrotvorskaya, M. V.; Fedorov, A. G.; Krivonogov, S. I.; Mateychenko, P. V.; Nashchekina, O. N.; Sipatov, A. Yu. Growth and structure of thermally evaporated $Bi_2Te_3$ thin films. Thin Solid Films 612, 128–134 (2016). DOI: 10.1016/j.tsf.2016.05.046.
25. Vovk, E. A.; Budnikov, A. T.; Nizhankovskiy, S. V.; Krivonogov, S. I.; Dobrotvorska, M. V.; Tkachenko, V. F.; Mateychenko, P. V. Structure and element composition of the nitride layer of $AlN/Al_2O_3$ templates obtained by the thermochemical nitridation of sapphire. Journal of Surface Investigation: X-ray, Synchrotron and Neutron Techniques 9, 1178–1183 (2015). DOI: 10.1134/S1027451015060221.
26. Smits, F. M. Measurement of Sheet Resistivities with the Four-Point Probe. *Bell System Technical Journal* **37**, 711–718 (1958). DOI: 10.1002/j.1538-7305.1958.tb03883.x.
27. Erbe, M.; Berrazouane, R.; Geyer, S.; Stasi, L.; van der Brugge, F.; Gras, G.; Schmidt, M.; Wieck, A. D.; Ludwig, A.; Bussières, F.; Warburton, R. J. *MoSi Superconducting Nanowire Single-Photon Detectors on GaAs for On-Chip Integration.* arXiv:2312.00528 (2023). DOI: 10.48550/arXiv.2312.00528.
28. Graybeal, J. M.; Beasley, M. R. Localization and interaction effects in ultrathin amorphous superconducting films. *Physical Review B* **29**, 4167(R) (1984). DOI: 10.1103/PhysRevB.29.4167.
29. Altshuler, B. L.; Aronov, A. G.; Lee, P. A. Interaction Effects in Disordered Fermi Systems in Two Dimensions. *Physical Review Letters* **44**, 1288–1291 (1980). DOI: 10.1103/PhysRevLett.44.1288.
30. Bergmann, G. Weak localization in thin films: a time-of-flight experiment with conduction electrons. *Physics Reports* **107**, 1–58 (1984). DOI: 10.1016/0370-1573(84)90103-0.
31. Buchstab, E. I.; Butenko, A. V.; Fogel, N. Ya.; Cherkasova, V. G.; Rosenbaum, R. L. Quantum corrections to the resistance of Mo/Si multilayers. *Physical Review B* **50**, 10063–10068 (1994). DOI: 10.1103/PhysRevB.50.10063.
32. Liu, Z.; Luo, B.; Hu, J.; Xing, C. Transport mechanism in amorphous molybdenum silicide thin films. *Journal of Physics and Chemistry of Solids* **149**, 109818 (2021). DOI: 10.1016/j.jpcs.2020.109818.

33. Mott, N. F. Conduction in non-crystalline materials. III. Localized states in a pseudogap and near extremities of conduction and valence bands. *Philosophical Magazine* **19**, 835–852 (1969). DOI: 10.1080/14786436908216338.
34. Efros, A. L.; Shklovskii, B. I. Coulomb gap and low temperature conductivity of disordered systems. *Journal of Physics C: Solid State Physics* **8**, L49–L51 (1975). DOI: 10.1088/0022-3719/8/4/003.